\documentclass[10pt,twocolumn,twoside]{IEEEtran}

\usepackage{graphicx}
\usepackage{epstopdf}
\usepackage{amsmath}
\usepackage{amssymb}
\usepackage{array}
\newcommand{\PreserveBackslash}[1]{\let\temp=\\#1\let\\=\temp}
\newcolumntype{C}[1]{>{\PreserveBackslash\centering}p{#1}}
\newcolumntype{R}[1]{>{\PreserveBackslash\raggedleft}p{#1}}
\newcolumntype{L}[1]{>{\PreserveBackslash\raggedright}p{#1}}
\usepackage[usenames]{color}
\usepackage{colortbl,booktabs}
\usepackage{tabularx}
\usepackage{multicol}
\usepackage{booktabs}
\usepackage[Symbol]{upgreek}
\usepackage{bm}
\usepackage{cite}
\usepackage{balance}
\usepackage[boxed]{algorithm2e}
\usepackage{mathrsfs}
\usepackage{url}
\usepackage{threeparttable}
\usepackage{amsmath}
\usepackage{dcolumn}
\usepackage{multirow}
\usepackage{booktabs}
\usepackage{amsfonts}

\DeclareMathOperator*{\argmax}{arg\,max}
\DeclareMathOperator*{\argmin}{arg\,min}
\DeclareMathOperator{\pr}{\text{Pr}}

\newcolumntype{d}[1]{D{.}{.}{#1}}

\begin{document}
\title{On the Performance of Channel Statistics-Based Codebook for Massive MIMO Channel Feedback}
\author{\IEEEauthorblockN{Wenqian Shen, Linglong Dai, Yu Zhang, Jianjun Li, and Zhaocheng Wang}
	
\thanks{Copyright (c) 2015 IEEE. Personal use of this material is permitted. However, permission to use this material for any other purposes must be obtained from the IEEE by sending a request to pubs-permissions@ieee.org.}

\thanks{W. Shen, L. Dai, Y. Zhang, and Z. Wang are with the Department of Electronic Engineering, Tsinghua University, Beijing 100084, China (e-mails: swq13@mails.tsinghua.edu.cn, daill@tsinghua.edu.cn, zhang-yu14@mails.tsinghua.edu.cn, and zcwang@tsinghua.edu.cn).}
\thanks{J. Li is with School of Electric and Information Engineering, Zhongyuan University of Technology, Zheng Zhou 450007, China (e-mail: jianjun.li@tsinghua.org.cn).}

\thanks{This work was supported by the National Key Basic Research Program of China (Grant No. 2013CB329203),  the National Natural Science Foundation of China (Grant Nos. 61571270 and 61571267),  the Beijing Natural Science Foundation (Grant No. 4142027), and the Foundation of Shenzhen government.}}

\maketitle
\vspace{-0mm}
\begin{abstract}
The channel feedback overhead for massive MIMO systems with a large number of base station (BS) antennas is very high, since the number of feedback bits of traditional codebooks scales linearly with the number of BS antennas. To reduce the feedback overhead, an effective codebook based on channel statistics has been designed, where the required number of feedback bits only scales linearly with the rank of the channel correlation matrix. However, this attractive conclusion was only proved under a particular channel assumption in the literature. To provide a rigorous theoretical proof under a general channel assumption, in this paper, we quantitatively analyze the performance of the channel statistics-based codebook. Specifically, we firstly introduce the rate gap between the ideal case of perfect channel state information at the transmitter and the practical case of limited channel feedback, where we find that the rate gap depends on the quantization error of the codebook. Then, we derive an upper bound of the quantization error, based on which we prove that the required number of feedback bits to ensure a constant rate gap only scales linearly with the rank of the channel correlation matrix. Finally, numerical results are provided to verify this conclusion. 
\end{abstract}

\IEEEpeerreviewmaketitle

\section{Introduction}
\IEEEPARstart{M}{assive} multiple-input multiple-output (MIMO) using hundreds of antennas at the base station (BS) is one of the key technologies for future 5G wireless communications \cite{SPM_FRusek_ScalingupMIMO}.
To achieve the expected high spectrum efficiency and energy efficiency of massive MIMO, accurate channel state information at the transmitter (CSIT) is crucial \cite{TVT_CQi_Pilotdesign,arxiv_CHuang_SMP}.
Utilizing the channel reciprocity, CSIT can be obtained from uplink channel estimation in time division duplexing (TDD) systems \cite{JSAC_XGao_EnergyEfficient}.
While most of existing works on massive MIMO consider TDD mode due to this reason,  frequency division duplexing (FDD) has many benefits over TDD and thus still dominates current cellular networks \cite{TVT_HXie_Unified}.
Therefore, it is important to study the CSIT acquisition problem for FDD massive MIMO systems.

However, the channel reciprocity does not exist in FDD systems, so accurate channel feedback from users to the BS is required.
Traditional codebooks for channel feedback such as Grassmannian codebook \cite{TIT_DJLove_Grassmannian},\cite{TIT_KKMukkavilli_Onbeamforming} and random vector quantization (RVQ) based codebook \cite{TWC_CKAu-yeung_RVQ},\cite{TIT_NJindal_MIMOBroadcast} have been extensively investigated.
For these codebooks, there is an important conclusion that, to maintain a constant capacity degradation due to channel quantization error,
the required number of feedback bits approximately scales linearly with the number of BS antennas \cite{TIT_NJindal_MIMOBroadcast}.
Thus, for massive MIMO with a large number of BS antennas, the channel feedback overhead will be overwhelming.
Therefore, several new codebooks have been proposed to reduce the channel feedback overhead.
 Based on the assumption of high correlation among BS antennas, an antenna-grouping based feedback scheme was proposed in \cite{TCOM_BLee_AntennaGrouping}. Another promising solution is the channel statistics-based codebook \cite{TVT_DJLove_LimitedFeedback} that is designed by multiplying each vector of an original codebook (e.g., RVQ-based codebook) by the channel correlation matrix.
	It has been verified through extensive simulations that the number of feedback bits required by this channel statistics-based codebook only scales linearly with the rank of the channel correlation matrix \cite{GLOBECOM_BClerckx_StaticticsCodebook}.
	As the rank of the channel correlation matrix is much smaller than the number of BS antennas in massive MIMO \cite{CL_WShen_LowRank}, the feedback overhead can be reduced.
	However, the authors of \cite{GLOBECOM_BClerckx_StaticticsCodebook} only provided a proof of this attractive conclusion under a particular channel assumption,
	where the singular values of the square root of the channel correlation matrix are the same except for the first one,
	while the rigorous proof under a general channel assumption has not been provided in the literature.

In this paper, we quantitatively analyze the performance of the channel statistics-based codebook under a general channel assumption\footnote{Simulation codes are provided to reproduce the results presented in this paper: \url{http://oa.ee.tsinghua.edu.cn/dailinglong/publications/publications.html}.}.
Specifically, we firstly introduce the rate gap between the ideal case of perfect CSIT and the practical case of limited channel feedback. We find that the rate gap depends on the quantization error of the codebook.
Although the quantization error is difficult to obtain, we are able to derive an upper bound of the quantization error by magnifying and shrinking the involved inequation.
After that, by substituting the upper bound of quantization error into the rate gap expression, we can obtain the upper bound of the rate gap.
Finally, we prove that the number of feedback bits required to ensure a constant rate gap scales linearly with the rank of the channel correlation matrix.
To the best of our knowledge, this work is the first one to provide a rigorous proof of this conclusion under a general channel assumption.

\textit{Notation}:
Lower-case and upper-case boldface letters denote vectors and matrices, respectively;
$(\cdot)^H$ and $(\cdot)^{-1}$ denote the conjugate transpose and inverse of a matrix, respectively;
$\mathbf{I}_K$ denotes the identity matrix of size $K\times K$;
$\text{E}\left[\cdot\right] $ denotes the expectation operator.
$\pr\left\lbrace A\right\rbrace $ denote the probability of $A$.
\section{System Model}
In this section, we firstly introduce the massive MIMO channel model. Then, we present the limited channel feedback. Finally, we review the per user rate.
\subsection{Massive MIMO Channel Model}\label{S2.1}
 In this paper, we consider a massive MIMO system with $M$ antennas at the BS and $K$ single-antenna users ($M\gg K$).
 The downlink channel vector $\mathbf{h}_k\in\mathbb{C}^{M\times 1}$ for the $k$-th user can be described as \cite{GLOBECOM_BClerckx_StaticticsCodebook}
 \begin{align}\label{eq_hk} 
 \vspace{-1mm}
 \mathbf{h}_k=\mathbf{R}_k^{1/2}\mathbf{h}_{w,k},
\vspace{-1mm}
 \end{align}
 where $\mathbf{R}_k^{1/2}\in\mathbb{C}^{M\times M}$ is the square root of the $k$-th user's channel correlation matrix\footnote{The channel correlation matrix $\mathbf{R}_k$ is closely related to the antenna spacing at the base station and the user location, which is the second-order channel statistics that can be usually assumed to be static due to the moderate user velocities \cite{TVT_DJLove_LimitedFeedback}.} $\mathbf{R}_k\in\mathbb{C}^{M\times M}$, and $\mathbf{h}_{w,k}\in\mathbb{C}^{M\times 1}$ is a vector whose elements are i.i.d. complex Gaussian distributed with zero mean and unit variance.
 Furthermore, $\mathbf{R}_k^{1/2}$ can be decomposed as $\mathbf{U}_k\mathbf{\Lambda}_k^{1/2}\mathbf{U}_k^H$,
 where $\mathbf{U}_k$ is an unitary matrix,
 and $\mathbf{\Lambda}_k^{1/2}$ is a diagonal matrix with the diagonal elements denoted by $\{\sigma_1,\sigma_2,\cdots,\sigma_r,0,\cdots\}$,
 where $r$ is the rank of correlation matrix.
 The concatenation of $K$ channel vectors can be denoted by $\mathbf{H}=[\mathbf{h}_1,\mathbf{h}_2,\cdots,\mathbf{h}_K] \in\mathbb{C}^{M\times K}$.
\subsection{Limited Channel Feedback}\label{S2.2}
Although the training overhead to obtain the downlink channel vector at the user side is increased in massive MIMO systems, there are some recently proposed effective downlink training methods \cite{CL_WShen_LowRank},\cite{TSP_ZGao_CommonSparsity} proposed with reduced training overhead. Thus, in this paper, each user is assumed to know its channel vector $\mathbf{h}_k$.
Such information is required at the BS via limited feedback channel.
Each user quantizes its channel to $B$ bits and then feeds them back to the BS.
Quantization is realized by using a quantization codebook, which is known to the BS and users.

For the traditional RVQ-based codebook \cite{TIT_NJindal_MIMOBroadcast} $\mathcal{W}=\{\mathbf{w}_1,\mathbf{w}_2,\cdots,\mathbf{w}_{2^B}\}$, the unit-norm column vector $\mathbf{w}_i\in\mathbb{C}^{M\times 1}$ is randomly generated by selecting vector independently from the uniform distribution on the complex unit sphere.
The required number of feedback bits scales linearly with the number of BS antennas to ensure a constant capacity degradation \cite{TIT_NJindal_MIMOBroadcast}.
Thus, the channel feedback overhead becomes overwhelming for massive MIMO with a large number of BS antennas.
In order to reduce the channel feedback overhead, a more effective codebook has been designed based on the channel statistics \cite{TVT_DJLove_LimitedFeedback}.
Specifically, the quantization vector $\mathbf{c}_{k,i}\in\mathbb{C}^{M\times 1}$ in the $k$-th users's codebook $\mathcal{C}_k$ can be obtained by multiplying the vector $\mathbf{w}_i$ by the square root of the channel correlation matrix $\mathbf{R}_k^{1/2}$, i.e., $\mathbf{c}_{k,i}=\mathbf{R}_k^{1/2}\mathbf{w}_i$.
Note that to ensure the unit-norm vector requirement of $\mathbf{c}_{k,i}\in\mathbb{C}^{M\times 1}$, $\mathbf{c}_{k,i}$ should be normalized as
\begin{align} \label{eq_ci}
\vspace{-1mm}
\mathbf{c}_{k,i}=\frac{\mathbf{R}_k^{1/2}\mathbf{w}_i}{\|\mathbf{R}_k^{1/2}\mathbf{w}_i\|}.
\vspace{-1mm}
\end{align}
The distribution of the resulted quantization vector $\mathbf{c}_{k,i}$ is closer to the distribution of actual channel vector $\mathbf{h}_k$ in (\ref{eq_hk}),
thus, the channel statistics-based codebook has better quantization performance \cite{TCOM_BLee_AntennaGrouping,TVT_DJLove_LimitedFeedback,GLOBECOM_BClerckx_StaticticsCodebook}.
Accordingly, we mainly consider the channel statistics-based codebook in this paper.

The $k$-th user quantizes its own channel $\mathbf{h}_k$ to a quantization vector $\mathbf{c}_{k,F_k}$ that is closest to $\mathbf{h}_k$, where ``closeness" is measured by the angle between two vectors. Thus, user $k$ computes the quantization index $F_k$ according to
\begin{align}
\vspace{-1mm}
F_k=\argmin_{i\in [1,2^B]}\sin^2(\measuredangle(\mathbf{h}_k,\mathbf{c}_{k,i}))=\argmax_{i\in [1,2^B]}|\mathbf{\tilde{h}}_k^H\mathbf{c}_{k,i}|^2,
\vspace{-3mm}
\end{align}
where the normalized channel vector $\mathbf{\tilde{h}}_k=\frac{\mathbf{h}_k}{\|\mathbf{h}_k\|}$ denotes the direction of channel vector.
Note that only the direction of channel vector is quantized, while the channel magnitude $\|\mathbf{h}_k\|$ is not quantized by using codebook $\mathbf{C}$.
Magnitude information is just a scalar value that can be easily fed back. In this paper, we focus on the quantization of channel direction.
After that, with the received index $F_k$, the BS can obtain the feedback channel vector $\hat{\mathbf{h}}_k=\|\mathbf{h}_k\|\mathbf{c}_{k,F_k}$.
The concatenation of the feedback channel vectors can be denoted as $\mathbf{\hat{H}}=[\mathbf{\hat{h}}_1,\mathbf{\hat{h}}_2,\cdots,\mathbf{\hat{h}}_K] \in\mathbb{C}^{M\times K}$.
\subsection{Downlink Precoding and Per User Rate}\label{S2.3}
We can utilize the widely used zero-forcing (ZF) precoding at the BS based on the feedback channel matrix $\mathbf{\hat{H}}$ to eliminate interferences among multiple users.
The transmitted signal $\mathbf{x}\in\mathbb{C}^{M\times 1}$ after precoding at the BS is given by
\begin{align}
\vspace{-1mm}
\mathbf{x}=\sqrt{\frac{\gamma}{K}}\mathbf{V}\mathbf{s},
\vspace{-1mm}
\end{align}
where $\gamma$ is the transmit power, $\mathbf{s}=[s_1,s_2,\cdots,s_K]\in\mathbb{C}^{K\times 1}$ is the symbol vector intended for $K$ users with the normalized power $\text{E}\left[|s_i|^2\right] =1$, and $\mathbf{V}=[\mathbf{v}_1,\mathbf{v}_2,\cdots,\mathbf{v}_K]\in\mathbb{C}^{M\times K}$ is the precoding matrix consisting of $K$ different $M$-dimensional unit-norm precoding vector $\mathbf{v}_i\in\mathbb{C}^{M\times 1}$.
We denote $\mathbf{U}=\mathbf{\hat{H}}(\mathbf{\hat{H}}^H\mathbf{\hat{H}})^{-1}$,
then the precoding vectors $\mathbf{v}_i$ can be obtained as the normalized $i$-th column of $\mathbf{U}$, i.e. $\mathbf{v}_i=\frac{\mathbf{U}(:,i)}{\|\mathbf{U}(:,i)\|}$.

 The received signal $y_k$ at the $k$-th user can be described as
 \begin{align}
 \vspace{-1mm}
y_k&=\mathbf{h}_k^H\mathbf{x}+n_k \\\nonumber
&=\sqrt{\frac{\gamma}{K}}\mathbf{h}_k^H\mathbf{v}_ks_k+\sqrt{\frac{\gamma}{K}}\sum_{i=1,i\neq k}^{K}\mathbf{h}_k^H\mathbf{v}_is_i+n_k,
\vspace{-1mm}
\end{align}
where $n_k$ is the complex Gaussian noise at the $k$-th user with zero mean and unit variance.
Thus, the signal-to-interference-plus-noise ratio (SINR) at the $k$-th user is \cite{TIT_NJindal_MIMOBroadcast}
 \begin{align}
\text{SINR}_k=\frac{\frac{\gamma}{K}|\mathbf{h}_k^H\mathbf{v}_k|^2}{1+\frac{\gamma}{K}\sum_{i=1,i\neq k}^{K}|\mathbf{h}_k^H\mathbf{v}_i|^2}.
\end{align}
Accordingly, the per user rate $R$ is
 \begin{align}
R=\text{E}\left[\log_2\left(1+\frac{\frac{\gamma}{K}|\mathbf{h}_k^H\mathbf{v}_k|^2}{1+\frac{\gamma}{K}\sum_{i=1,i\neq k}^{K}|\mathbf{h}_k^H\mathbf{v}_i|^2}\right) \right] .
\end{align}

Clearly, the per user rate depends on the precoding matrix $\mathbf{V}$,
which is significantly affected by the quality of feedback channel $\mathbf{\hat{H}}$.
In the next section,
we will analyze the per user rate when the channel statistics-based codebook is applied.

\section{Performance Analysis}\label{S3}
In this section, we firstly compute the rate gap.
Then, we analyze the quantization error of the channel statistics-based codebook.
Finally, we derive an upper bound of the required feedback bits to ensure a constant rate gap.

\subsection{Rate Gap}\label{S3.1}
Providing the ideal case of perfect CSIT at the BS, i.e., $\mathbf{\hat{H}}=\mathbf{H}$,
the ZF precoding vector $\mathbf{v}_{\text{ideal},i}$ is obtained as the normalized $i$-th column of $\mathbf{H}(\mathbf{H}^H\mathbf{H})^{-1}$.
Therefore, we can obtain the ideal per user rate as
 \begin{align}
R_\text{ideal}=\text{E}\left[\log_2\left(1+\frac{\gamma}{K}|\mathbf{h}_k^H\mathbf{v}_{\text{ideal},k}|^2\right) \right] .
\end{align}

However, in the practical case of limited channel feedback, the BS can only obtain the feedback channel $\mathbf{\hat{H}}$ using the channel statistics-based codebook.
The ZF precoding is performed based on $\mathbf{\hat{H}}$,
and the precoding vector $\mathbf{v}_i$ is obtained as the normalized $i$-th column of $\mathbf{\hat{H}}(\mathbf{\hat{H}}^H\mathbf{\hat{H}})^{-1}$.
Thus, the inter-user interference $|\mathbf{h}_k^H\mathbf{v}_i|\neq 0$, which degrades the per user rate as
 \begin{align}
R_\text{practical}=\text{E}\left[\log_2\left(1+\frac{\frac{\gamma}{K}|\mathbf{h}_k^H\mathbf{v}_k|^2}{1+\frac{\gamma}{K}\sum_{i=1,i\neq k}^{K}|\mathbf{h}_k^H\mathbf{v}_i|^2}\right) \right] .
\end{align}

We define the rate gap $\Delta R(\gamma)$ as the difference between per user rate achieved by ideal CSIT and limited channel feedback using the channel statistics-based codebook:
\begin{align}
 \vspace{-1mm}
\Delta R(\gamma) &= R_\text{ideal}-R_\text{practical}.
 \vspace{-1mm}
\end{align}
Following the results from \cite{TIT_NJindal_MIMOBroadcast, TCOM_BLee_AntennaGrouping} and using Jensen's inequality, the rate gap $\Delta R(\gamma)$ can be upper bounded as:
\begin{align}\label{delatR1}
 \vspace{-1mm}
\Delta R(\gamma)
&\leq \log_2\left(1+\frac{\gamma}{K}(K-1)\text{E}\left[|\mathbf{h}_k^H\mathbf{v}_i|^2\right] \right) ,
 \vspace{-1mm}
\end{align}
where the multi-user interference $\text{E}\left[|\mathbf{h}_k^H\mathbf{v}_i|^2\right] $ can be upper bounded in \textit{Lemma 1}.

\textbf{\textit{Lemma 1}}:
The upper bound of multi-user interference $\text{E}\left[|\mathbf{h}_k^H\mathbf{v}_i|^2\right] $ depends on the channel quantization error $\text{E}\left[\sin^2(\measuredangle(\tilde{\mathbf{h}}_k,\hat{\mathbf{h}}_k))\right] $, i.e.,
\begin{align}\label{EhHvj}
 \vspace{-1mm}
\text{E}\left[|\mathbf{h}_k^H\mathbf{v}_i|^2\right] \leq
\text{E}\left[\|\mathbf{h}_k\|^2\right] \text{E}\left[\sin^2(\measuredangle(\tilde{\mathbf{h}}_k,\hat{\mathbf{h}}_k))\right] .
 \vspace{-1mm}
\end{align}
\begin{IEEEproof}
Denote the quantization error $X=\sin^2(\measuredangle(\tilde{\mathbf{h}}_k,\hat{\mathbf{h}}_k))=1-|\mathbf{\tilde{h}}_k^H\mathbf{c}_{k,F_k}|^2$.
Since the size of codebook is limited, $X\neq 0$.
Thus, the normalized channel vector $\mathbf{\tilde{h}}_k$ can be decomposed along two orthogonal direction,
one is the direction of quantization vector $\mathbf{c}_{k,F_k}$, and the other is in the nullspace of $\mathbf{c}_{k,F_k}$.
Mathmatically,
 \begin{align}\label{eq_decomp}
\tilde{\mathbf{h}}_k=\sqrt{1-X}\mathbf{c}_{k,F_k}+\sqrt{X}\mathbf{s},
\end{align}
where $\mathbf{s}$ is an unit vector distributed in the null space of $\mathbf{c}_{k,F_k}$ \cite{TIT_NJindal_MIMOBroadcast}.
Combining $\mathbf{h}_k=\|\mathbf{h}_k\|\tilde{\mathbf{h}}_k$ and (\ref{eq_decomp}), we have
\begin{align}
 \vspace{-1mm}
|\mathbf{h}_k^H\mathbf{v}_i|^2=\|\mathbf{h}_k\|^2\left( (1-X)|\mathbf{c}_{k,F_k}^H\mathbf{v}_i|^2+X|\mathbf{s}^H\mathbf{v}_i|^2\right) .
 \vspace{-1mm}
\end{align}
Since the precoding vector $\mathbf{v}_i$ is obtained as the normalized $i$-th column of $\mathbf{\hat{H}}(\mathbf{\hat{H}}^H\mathbf{\hat{H}})^{-1}$,
$\mathbf{v}_i$ is orthogonal to the $k$-th user's feedback channel vector $\mathbf{\hat{h}}_k=\|\mathbf{h}_k\|\mathbf{c}_{k,F_k}$, i.e., $\mathbf{c}_{k,F_k}^H\mathbf{v}_i=0$.
Thus, we have $|\mathbf{h}_k^H\mathbf{v}_i|^2=\|\mathbf{h}_k\|^2X|\mathbf{s}^H\mathbf{v}_i|^2$.
Since $|\mathbf{s}^H\mathbf{v}_i|^2=\cos^2(\measuredangle(\mathbf{s},\mathbf{v}_i)) \leq 1$, we have
\begin{align} \label{eq_hkvi}
|\mathbf{h}_k^H\mathbf{v}_i|^2{\leq}\|\mathbf{h}_k\|^2X.
\end{align}
As the norm of a vector is independent of its direction \cite{TIT_NJindal_MIMOBroadcast}, $\|\mathbf{h}_k\|^2$ is independent of $X$.
Therefore, we obtain (\ref{EhHvj})
\end{IEEEproof}

Combining (\ref{delatR1}) and  \textit{Lemma 1}, we can further obtain
\begin{align}\label{eq_Rg_upb}
\Delta R(\gamma)\!\leq\! \log_2\left(1\!+\!\frac{\gamma}{K}(K\!-\!1)\text{E}\!\left[\|\mathbf{h}_k\|^2\right] \text{E}\!\left[\sin^2(\measuredangle(\tilde{\mathbf{h}}_k,\hat{\mathbf{h}}_k))\right] \right) .
\end{align}
\subsection{Quantization Error}\label{S3.2}
In this subsection, we discuss the quantization error $\text{E}\left[\sin^2(\measuredangle(\tilde{\mathbf{h}}_k,\hat{\mathbf{h}}_k))\right] $ in (\ref{eq_Rg_upb}) when the channel statistics-based codebook is considered.
For the rest of this paper, we omit the subscript $k$ for simplicity but without loss of generality.

\textbf{ \textit{Lemma 2}}:
The quantization error $\text{E}\left[\sin^2(\measuredangle(\tilde{\mathbf{h}},\hat{\mathbf{h}}))\right] $ of $\tilde{\mathbf{h}}$ can be upper bounded as
\begin{align}\label{eq_qua_err_upb}
\text{E}\left[\sin^2(\measuredangle(\tilde{\mathbf{h}},\hat{\mathbf{h}}))\right] <2^{-\frac{B}{r-1}}.
\end{align}

\begin{IEEEproof}
Denote that $Z=\cos^2(\measuredangle(\tilde{\mathbf{h}},\hat{\mathbf{h}}))=|\mathbf{\tilde{h}}^H\mathbf{c}_{F}|^2$.
Since $F=\argmax_{i\in[1,2^B]}|\mathbf{\tilde{h}}^H\mathbf{c}_{i}|^2$,
we have $\pr\left\lbrace |\mathbf{\tilde{h}}^H\mathbf{c}_{F}|^2<z\right\rbrace =\pr\left\lbrace |\mathbf{\tilde{h}}^H\mathbf{c}_1|^2<z,|\mathbf{\tilde{h}}^H\mathbf{c}_2|^2<z,\cdots\mathbf{\tilde{h}}^H\mathbf{c}_{2^B}|^2<z\right\rbrace \overset{(a)}{=}\pr\left\lbrace |\mathbf{\tilde{h}}^H\mathbf{c}_{i}|^2<z\right\rbrace ^{2^B}$,
where (a) is true due to the independence among $\{|\mathbf{\tilde{h}}^H\mathbf{c}_{i}|^2\}_{i=1}^{2^B}$ \cite{TIT_NJindal_MIMOBroadcast}.
Therefore,
\begin{align}\label{eq_prZ}
\pr\left\lbrace Z<z\right\rbrace =\pr\left\lbrace |\mathbf{\tilde{h}}^H\mathbf{c}_{i}|^2<z\right\rbrace ^{2^B}.
\end{align}
Using (\ref{eq_hk}) and (\ref{eq_ci}), we can rewrite $|\mathbf{\tilde{h}}^H\mathbf{c}_{i}|^2$ as
\begin{align}
|\mathbf{\tilde{h}}^H\mathbf{c}_{i}|^2&=\frac{|(\mathbf{R}^{1/2}\mathbf{h}_w)^H\mathbf{R}^{1/2}\mathbf{w}_i|^2}{\|\mathbf{R}^{1/2}\mathbf{h}_w\|^2\|\mathbf{R}^{1/2}\mathbf{w}_i\|^2}\\\nonumber
&\overset{(a)}{=}\frac{|\mathbf{h}_w^H\mathbf{U}(\mathbf{\Lambda}^{1/2})^H\mathbf{\Lambda}^{1/2}\mathbf{U}^H\mathbf{w}_i|^2}{\|\mathbf{R}^{1/2}\mathbf{h}_w\|^2\|\mathbf{R}^{1/2}\mathbf{w}_i\|^2},
\end{align}
where (a) is obtained by using $\mathbf{U}^H\mathbf{U}=\mathbf{I}_M$.
Since the unitary matrix $\mathbf{U}$ does not change the distribution of an isotropically distributed vector,
$\mathbf{h}_w^H\overset{d}{=}\mathbf{h}_w^H\mathbf{U}$ and $\mathbf{w}_i\overset{d}{=}\mathbf{U}^H\mathbf{w}_i$, where $\overset{d}{=}$ denotes the equality in terms of distribution.
Thus,
\begin{align}
|\mathbf{\tilde{h}}^H\mathbf{c}_{i}|^2\overset{d}{=}\frac{|\mathbf{h}_w^H(\mathbf{\Lambda}^{1/2}_k)^H\mathbf{\Lambda}^{1/2}_k\mathbf{w}_i|^2}{\|\mathbf{\Lambda}^{1/2}_k\mathbf{h}_w\|^2\|\mathbf{\Lambda}^{1/2}_k\mathbf{w}_i\|^2},
\end{align}
where the denominator is accordingly changed to ensure the unit-norm requirement of $\mathbf{\tilde{h}}$ and $\mathbf{c}_{i}$.
Therefore, we have
\begin{align}\label{hHci}
\pr\left\lbrace|\mathbf{\tilde{h}}^H\mathbf{c}_{i}|^2<z\right\rbrace =\pr\left\lbrace\frac{|\mathbf{h}_w^H(\mathbf{\Lambda}^{1/2})^H\mathbf{\Lambda}^{1/2}\mathbf{w}_i|^2}{\|\mathbf{\Lambda}^{1/2}\mathbf{h}_w\|^2\|\mathbf{\Lambda}^{1/2}\mathbf{w}_i\|^2}<z\right\rbrace .
\end{align}

Since ${\Lambda}^{1/2}$ is an diagonal matrix with $r$ non-zero diagonal elements $\{\sigma_1, \sigma_2,\cdots,\sigma_r\}$, we rewrite (\ref{hHci}) as
\begin{align}\label{eq_hHciRewrite}
\pr\left\lbrace|\mathbf{\tilde{h}}^H\mathbf{c}_{i}|^2<z\right\rbrace =\pr\left\lbrace\frac{|\mathbf{g}^H\mathbf{\Gamma}^H\mathbf{\Gamma}\mathbf{v}|^2}{\|\mathbf{\Gamma}\mathbf{g}\|^2\|\mathbf{\Gamma}\mathbf{v}\|^2}<z\right\rbrace ,
\end{align}
where $\mathbf{g}\in\mathbb{C}^{r\times1}$ with $\mathbf{g}(j)=\mathbf{h}_w(j)$, $\mathbf{v}\in\mathbb{C}^{r\times1}$ with $\mathbf{v}(j)=\mathbf{w}_i(j)$, and $\mathbf{\Gamma}\in\mathbb{C}^{r\times r}$ with $\mathbf{\Gamma}(j,j)=\mathbf{\Lambda}^{1/2}(j,j), j=1,2,\cdots ,r$.
The non-diagonal elements $\mathbf{\Gamma}(i,j)=0$.
Actually, $\mathbf{g}$ and $\mathbf{v}$ are random vectors in the $r$-dimensional hyper-sphere,
while $\mathbf{\Gamma}\mathbf{g}$ and $\mathbf{\Gamma}\mathbf{v}$ are randomly distributed in the $r$-dimensional hyper-ellipse, which is obtained by stretching the hyper-sphere according to the diagonal elements of $\mathbf{\Gamma}$.
As we known, $\frac{|\mathbf{g}^H\mathbf{v}|^2}{\|\mathbf{g}\|^2\|\mathbf{v}\|^2}$ is the squared cosine of the angle between two vectors $\mathbf{g}$ and $\mathbf{v}$ in hyper-sphere, whose cumulative distribution function (CDF) is given by $\pr\left\lbrace\frac{|\mathbf{g}^H\mathbf{v}|^2}{\|\mathbf{g}\|^2\|\mathbf{v}\|^2}\leq z\right\rbrace =1-(1-z)^{r-1}$ for $z\in [0,1]$ \cite{TWC_CKAu-yeung_RVQ}.
In (\ref{eq_hHciRewrite}), $\frac{|\mathbf{g}^H\mathbf{\Gamma}^H\mathbf{\Gamma}\mathbf{v}|^2}{\|\mathbf{\Gamma}\mathbf{g}\|^2\|\mathbf{\Gamma}\mathbf{v}\|^2}$ is the squared cosine of the angle between two vectors $\mathbf{\Gamma}\mathbf{g}$ and $\mathbf{\Gamma}\mathbf{v}$ in the hyper-ellipse, whose CDF is very difficult be obtained.
However, we can prove in the \textbf{Appendix A} that $\pr\left\lbrace\frac{|\mathbf{g}^H\mathbf{\Gamma}^H\mathbf{\Gamma}\mathbf{v}|^2}{\|\mathbf{\Gamma}\mathbf{g}\|^2\|\mathbf{\Gamma}\mathbf{v}\|^2}\leq z\right\rbrace <\pr\left\lbrace\frac{|\mathbf{g}^H\mathbf{v}|^2}{\|\mathbf{g}\|^2\|\mathbf{v}\|^2}\leq z\right\rbrace $.
Thus, we have
\begin{align}\label{eqCDFup}
\pr\left\lbrace\frac{|\mathbf{g}^H\mathbf{\Gamma}^H\mathbf{\Gamma}\mathbf{v}|^2}{\|\mathbf{\Gamma}\mathbf{g}\|^2\|\mathbf{\Gamma}\mathbf{v}\|^2}<z\right\rbrace <1-(1-z)^{r-1}.
\end{align}

Then, by combining (\ref{eq_prZ}), (\ref{eq_hHciRewrite}), and (\ref{eqCDFup}), we can obtain the upper bound of $Z$'s CDF as
\begin{align} \label{eq_Z_CDF}
\pr\left\lbrace Z<z\right\rbrace <(1-(1-z)^{r-1})^{2^B}.
 \end{align}
Utilizing the fact that $\text{E}\left[Z\right] =\int_0^1\pr\left\lbrace Z\geq z\right\rbrace $, the expectation $\text{E}\left[Z\right] $ can be expressed as
\begin{align} \label{eq_Z_jifen}
\text{E}\left[Z\right] &=\int_0^1(1-\pr\left\lbrace  Z<z\right\rbrace )dz=1-\int_0^1\pr\left\lbrace Z<z\right\rbrace dz.
\end{align}
Combining (\ref{eq_Z_CDF}) and (\ref{eq_Z_jifen}), we have
\begin{align}
\text{E}\left[Z\right] &>1-\int_0^1 (1-(1-z)^{r-1})^{2^B}dz\\\nonumber
&\overset{(a)}{=}1-\int_0^1 (1-s^{r-1})^{2^B}ds\\\nonumber
&\overset{(b)}{=}1-2^B\beta\left(2^B,\frac{r}{r-1}\right)\overset{(c)}{\geq}1-2^{-\frac{B}{r-1}},
\end{align}
where (a) is obtained by setting $s=1-z$, (b) and (c) are obtained from \cite[Appendix I and II]{TIT_NJindal_MIMOBroadcast}, respectively.
Finally, we can obtain the upper bound of quantization error as
\begin{align}\label{eq_sin2h}
 \vspace{-1mm}
\text{E}\left[\sin^2(\measuredangle(\tilde{\mathbf{h}},\hat{\mathbf{h}}))\right] =1-\text{E}\left[Z\right] <2^{-\frac{B}{r-1}}.
 \vspace{-2mm}
\end{align}
\end{IEEEproof}
\subsection{Feedback Bits}\label{S3.3}
In this subsection, we discuss the required number of feedback bits $B$ to ensure a constant rate gap $\Delta R(\gamma)$.
By combining (\ref{eq_Rg_upb}) and (\ref{eq_sin2h}), we can easily obtain
\begin{align}\label{delatR_UpBound}
\Delta R(\gamma)&\leq\log_2\left(1+\frac{\gamma}{K}(K-1)\text{E}\left[\|\mathbf{h}\|^2\right] 2^{-\frac{B}{r-1}}\right) .
\end{align}
To ensure the rate gap $\Delta R(\gamma) \leq \log_2\left(b\right) $, based on (\ref{delatR_UpBound}) and letting $\log_2\left(1+\frac{\gamma}{K}(K-1)\text{E}\left[\|\mathbf{h}\|^2\right] 2^{-\frac{B}{r-1}}\right) \leq \log_2\left(b\right) $ bps/Hz, we have
\begin{align}\label{eq_B}
B \geq \frac{r-1}{3}\text{SNR}+(r-1)\log_2\left(\frac{K-1}{b-1}\right) ,
\end{align}
where the signal-to-noise-ratio (SNR) at the receiver is defined as $\text{SNR}=10\log_{10}\frac{\gamma}{K}\text{E}\left[\|\mathbf{h}\|^2\right] $.
We can observe that required number of feedback bits only scales linearly with the rank of correlation matrix $r$ when $\text{SNR}$ increases.
\section{Simulation Verification}
\begin{figure} [t]
	\vspace{-0mm}
	\center{\includegraphics[width=0.4\textwidth]{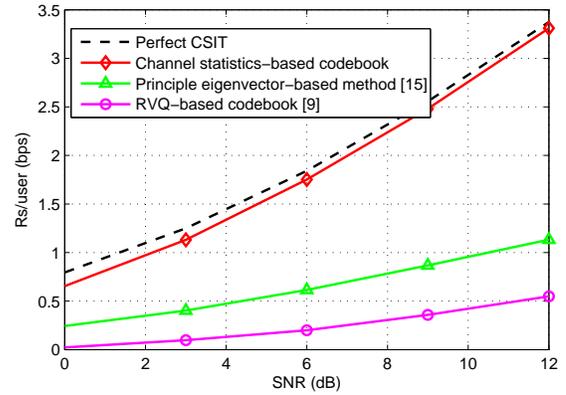}}
	\vspace{-2mm}
	\caption{Per user rate against the SNR at the receiver.}
	\label{Fig1}
\end{figure}

\begin{figure}[h]
\vspace{-0mm}
\center{\includegraphics[width=0.4\textwidth]{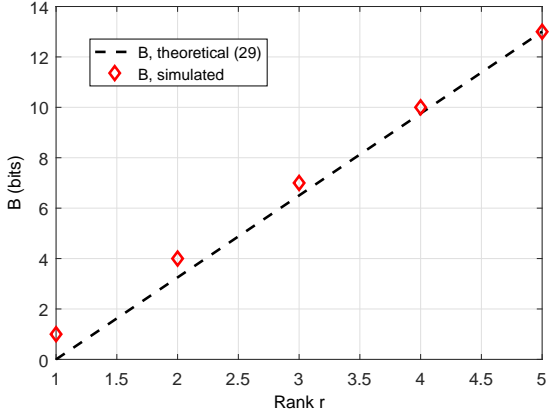}}
\vspace{-2mm}
\caption{The number of feedback bits against the rank of correlation matrix.}
\label{Fig2}
\end{figure}

In this section, simulation results are provided to verify the derived theoretical result.
The simulation setup is as follows:
the number of BS antennas, the number of users, and the rank of the channel correlation matrix is set as $(M, K, r)=(64, 10, 4)$;
the square root of the channel correlation matrix is $\mathbf{R}_k^{1/2}=\mathbf{U}_k\mathbf{\Lambda}_k^{1/2}\mathbf{U}_k^H$,
where $\mathbf{U}_k$ is an random unitary matrix, and $\mathbf{\Lambda}_k^{1/2}$ is a diagonal matrix with $r$ non-zero diagonal elements \cite{GLOBECOM_BClerckx_StaticticsCodebook};
the feedback bits $B = \frac{r-1}{3}\text{SNR}+3.17$ for Fig. \ref{Fig1} and $\text{SNR}=6$ dB for Fig. \ref{Fig2}.

Fig. \ref{Fig1} shows the per user rate against SNR.
We can observe from Fig. \ref{Fig1} that the rate gap between the ideal case of perfect CSIT and the practical case using the channel statistics-based codebook to realize CSIT remains constant when SNR increases.
Such simulated result is consistent with our theoretical analysis in Section III.
On the other hand, we can also find that the rate gap between the ideal case of perfect CSIT and the practical case using RVQ-based codebook to realize CSIT becomes greater when SNR increases, since the required number of feedback bits for RVQ-based codebook should scale linearly with the number of BS antennas, which is much larger than the rank of the channel correlation matrix. 
Moreover, we also show the per user rate of the principle eigenvector-based method \cite{TSP_SZhou_Eigenbeamforming} for comparison, which is overperformed by the channel statistics-based codebook.

Fig. \ref{Fig2} shows the required number of feedback bits to limit the rate gap between the ideal case of perfect CSIT and the practical case of using the channel statistics-based codebook within 0.09 bps/Hz. We can observe that the required number of feedback bits $B$ scales linearly with the rank of the channel correlation matrix $r$, which is consistent with the theoretical result (\ref{eq_B}) shown by the dash curve in Fig. \ref{Fig2}.

\section{Conclusions}
It has been shown that the number of feedback bits scales linearly with the rank of the channel correlation matrix  for the channel statistics-based codebook.
However, this attractive conclusion was only proved under a particular channel assumption in the literature.
To this end, in this paper we provided a rigorous proof under a general channel assumption for the first time.
At first, we found that the rate gap between the ideal case of perfect CSIT and using the channel stastics-based codebook depends on the channel quantization error.
Then, we derived an upper bound of the quantization error, based on which we proved that the required number of feedback bits scales linearly with the rank of the channel correlation matrix.
Moreover, the proof techniques in this paper can be used to analyze the performance of other codebooks.
\section{Appendix A}
The CDF of the squared cosine of the angle between two vectors $\mathbf{\Gamma}\mathbf{g}$ and $\mathbf{\Gamma}\mathbf{v}$ in the hyper-ellipse is upper bounded as:
\begin{align} \label{eq_CDF_com1}
\pr\left\lbrace\frac{|\mathbf{g}^H\mathbf{\Gamma}^H\mathbf{\Gamma}\mathbf{v}|^2}{\|\mathbf{\Gamma}\mathbf{g}\|^2\|\mathbf{\Gamma}\mathbf{v}\|^2}\leq z\right\rbrace <\pr\left\lbrace\frac{|\mathbf{g}^H\mathbf{v}|^2}{\|\mathbf{g}\|^2\|\mathbf{v}\|^2}\leq z\right\rbrace .
\end{align}

\begin{IEEEproof}
Firstly, we consider the squared cosine of the angle $\frac{|\mathbf{g}^H\mathbf{v}|^2}{\|\mathbf{g}\|^2\|\mathbf{v}\|^2}$ between two isotropic vectors $\mathbf{g}$ and $\mathbf{v}$ uniformly distributed in a hyper-sphere.
For $z\in [0,1]$, the CDF of $\frac{|\mathbf{g}^H\mathbf{v}|^2}{\|\mathbf{g}\|^2\|\mathbf{v}\|^2}$ is given by \cite{TWC_CKAu-yeung_RVQ}
\begin{align} \label{eq_CDF}
\pr\left\lbrace\frac{|\mathbf{g}^H\mathbf{v}|^2}{\|\mathbf{g}\|^2\|\mathbf{v}\|^2}\leq z\right\rbrace =1-(1-z)^{r-1}.
\end{align}
For $\frac{|\mathbf{g}^H\mathbf{\Gamma}^H\mathbf{\Gamma}\mathbf{v}|^2}{\|\mathbf{\Gamma}\mathbf{g}\|^2\|\mathbf{\Gamma}\mathbf{v}\|^2}$ in (\ref{eq_CDF_com1}),
the elements of $\mathbf{g}$ and $\mathbf{v}$ are multiplied by the diagonal elements $\{\sigma_1,\sigma_2,\cdots,\sigma_r\}$ of $\mathbf{\Gamma}$.
The uniformly distributed vectors $\mathbf{g}$ and $\mathbf{v}$ in hyper-sphere are projected onto non-uniformly distributed vectors $\mathbf{\Gamma}\mathbf{g}$ and $\mathbf{\Gamma}\mathbf{v}$ in hyper-ellipse,
which is generated by stretching the hyper-sphere.
That means the angle between two vectors in hyper-ellipse tends to be smaller,
i.e., the squared cosine of the angle between two vectors in hyper-ellipse tends to be larger.
Thus, the CDF of $\frac{|\mathbf{g}^H\mathbf{\Gamma}^H\mathbf{\Gamma}\mathbf{v}|^2}{\|\mathbf{\Gamma}\mathbf{g}\|^2\|\mathbf{\Gamma}\mathbf{v}\|^2}$ is smaller than that of $\frac{|\mathbf{g}^H\mathbf{v}|^2}{\|\mathbf{g}\|^2\|\mathbf{v}\|^2}$, i.e., we have (\ref{eq_CDF_com1}).

Now we give an intuitive example of an extreme case where $\sigma_1=\sigma_2=,\cdots,=\sigma_{r-1}$ and $\sigma_r\ll\sigma_2$.
The $r$-dimensional hyper-ellipse is generated by compress the last dimension of a $r$-dimensional hyper-sphere.
Thus, the vectors in hyper-sphere are projected onto a compressed hyper-sphere, i.e., the hyper-ellipse.
The angle between two vectors in hyper-ellipse becomes smaller.
Mathematically,
the numerator of the squared cosine can be approximately expressed as
\begin{align}
|\mathbf{g}^H\mathbf{\Gamma}^H\mathbf{\Gamma}\mathbf{v}|^2\approx |\sigma_1|^4|\Sigma_{i=1}^{r-1}g_i^*v_i|^2.
\end{align}
The dominator of the squared cosine can be approximately expressed as 
$\|\mathbf{\Gamma}\mathbf{g}\|^2\|\mathbf{\Gamma}\mathbf{v}\|^2\approx |\sigma_1|^4(\Sigma_{i=1}^{r-1}|g_i|^2)(\Sigma_{i=1}^{r-1}|v_i|^2)$.
Thus, we have 
$\frac{|\mathbf{g}^H\mathbf{\Gamma}^H\mathbf{\Gamma}\mathbf{v}|^2}{\|\mathbf{\Gamma}\mathbf{g}\|^2\|\mathbf{\Gamma}\mathbf{v}\|^2}\approx\frac{ |\Sigma_{i=1}^{r-1}g_i^*v_i|^2}{(\Sigma_{i=1}^{r-1}|g_i|^2)(\Sigma_{i=1}^{r-1}|v_i|^2)}$.
Considering (\ref{eq_CDF}), we can get
\begin{align}
\pr\left\lbrace\frac{|\mathbf{g}^H\mathbf{\Gamma}^H\mathbf{\Gamma}\mathbf{v}|^2}{\|\mathbf{\Gamma}\mathbf{g}\|^2\|\mathbf{\Gamma}\mathbf{v}\|^2}\leq z\right\rbrace \approx 1-(1-z)^{r-2}.
\end{align}
Utilizing the fact that $1-(1-z)^{r-2}<1-(1-z)^{r-1}$, we can obtain (\ref{eq_CDF_com1}).
\end{IEEEproof}


\bibliographystyle{IEEEtran}
\bibliography{IEEEabrv,Gao1Ref}

\end{document}